\documentclass[aip,apl,twocolumn,reprint,amsmath,amssymb,superscriptaddress]{revtex4-1}
\usepackage{graphicx}
\usepackage{epstopdf}
\usepackage{dcolumn}
\usepackage{bm}

\begin{document}

\preprint{final draft}

\title{Lateral 2D superlattices in GaAs heterostructures with independent control of carrier density and modulation potential}

\author{D. Q. Wang}
\email{qingwen.wang@unsw.edu.au}
\affiliation{School of Physics, University of New South Wales, Sydney NSW 2052, Australia}
\affiliation{Australian Research Council Centre of Excellence in Future Low-Energy Electronics Technologies, University of New South Wales, Sydney NSW 2052, Australia}
\author{D. Reuter}
\thanks{Now at Paderborn University, Department of Physics, Warburger Straße 100, 33098 Paderborn, Germany}
\author{A. D. Wieck}
\affiliation{Angewandte Festkörperphysik, Ruhr-Universität Bochum, D-44780 Bochum, Germany}
\author{A. R. Hamilton}
\affiliation{School of Physics, University of New South Wales, Sydney NSW 2052, Australia}
\affiliation{Australian Research Council Centre of Excellence in Future Low-Energy Electronics Technologies, University of New South Wales, Sydney NSW 2052, Australia}
\author{O. Klochan}
\affiliation{School of Physics, University of New South Wales, Sydney NSW 2052, Australia}
\affiliation{Australian Research Council Centre of Excellence in Future Low-Energy Electronics Technologies, University of New South Wales, Sydney NSW 2052, Australia}
\affiliation{School of Science, University of New South Wales, Canberra ACT 2612, Australia}

\date{\today}
\begin{abstract}
We present a new double-layer design for 2D surface superlattice systems in GaAs-AlGaAs heterostructures. Unlike previous studies, our device (1) uses an in-situ gate, which allows very short period superlattice in high mobility, shallow heterostructures; (2) enables independent control of the carrier density and the superlattice modulation potential amplitude over a wide range. We characterise this device design using low-temperature magneto-transport measurements and show that the fabrication process caused minimal damage to the system. We demonstrate the tuning of potential modulation from weak (much smaller than Fermi energy) to strong (larger than the Fermi energy) regimes. 
\end{abstract}

\maketitle
Moir\'{e} superlattices in two-dimensional (2D) materials are currently attracting significant attention as they enable the modification of bandstructure and thus the manipulation of electrical properties, beyond the limitations of natural crystals.
These artificially engineered superlattices exhibit a range of properties including unconventional superconductivity~\cite{Cao18_1, Yankowitz19, Lu19}, strongly correlated states~\cite{Cao18_2, Chen19, Lu19} , indirect valley excitons~\cite{Seyler19} and anomalous Hall effect~\cite{Serlin20}.
Artificial superlattices can also be applied to the 2D electron gas formed in conventional semiconductor heterostructures where the existing electronic properties can be altered by lithographical patterning~\cite{Park09, Givertini09}. Compared to Moir\'{e} systems, heterostructure based superlattices have the advantage of controllable lattice parameters: the superlattice can be made with any lattice symmetry, and a wide range of lattice constants. Lattice defects can be deliberately introduced by changing the lithography. 
Moreover, by tuning the properties of the original semiconductor system, such as spin-orbit coupling, we have the possibility of realising a topological insulator in those artificially engineered systems~\cite{Sushkov13}.

The difficulty of creating artificial bandstructures in a semiconductor based superlattice system lies in fabricating a low-disorder and small-period superlattice while maintaining the sample mobility at the same time. 
The artificial bandgap created by a periodic modulation potential depends on two parameters of the system. One is the lattice constant $a$, to which the size of the mini-bandgap is inversely proportional~\cite{Givertini09, Tkachenko15}. This means a small lattice constant $a\lesssim 150nm$ is required to overcome thermal and disorder-induced broadening at cryogenic temperatures. 
The other is the strength of the modulation potential $U$, which decays exponentially as a function of the distance from the gate to the 2DEG~\cite{Tkachenko15}. 
Since the 2DEG in conventional semiconductor heterostructures is generally buried tens of nanometers from the surface, a very shallow heterostructure is often required. However, compared to deep heterostructures, these shallow ones are inevitably more prone to damage caused by fabrication processes, and also scattering from surface states~\cite{Wang13}.

Previous studies of 2D superlattice modulated semiconductor 2DEGs have mainly focused on the GaAs material system~\cite{Weiss91_2, Weiss91, Geisler04, Kato12, Singha11, Wang18}.
Early work employing deep etching to remove part of the heterostructure revealed so-called ``Weiss" oscillations in the magnetoresistances due to scattering from the periodic potential~\cite{Weiss91_2, Weiss91}. 
However, these studies were mainly performed on modulation-doped structures with a comparatively deep 2DEG and large superlattice periods, making them unsuitable for artificial bandstructure investigations.
Recently smaller-period superlattices were achieved using similar deep-etching approach~\cite{Singha11, Wang18} but since the electron mobility in these devices is greatly reduced due to the etching, observing artificial bandstructures in transport measurements was not possible. 
Another method to create a lateral superlattice is using the strain caused by patterned resist on wafer surface through the piezo-electric effect~\cite{Geisler04, Kato12}. In this approach, damage to the heterostructure is reduced but the modulation potential is too weak to resolve artificial bandstructures~\cite{Givertini09, Tkachenko15}. 
Therefore, how to find the suitable modulation strength while maintaining the quality of the heterostructure remains a key challenge. 
Moreover, it is hard to tune the Fermi energy over a wide range in the modulation-doped heterostructures used to date. In contrast, the carrier density can be widely tuned in Moir\'{e} superlattices of 2D materials, making it much easier to probe the artificial bandstructure effects.

In this paper, we present a double-layer design that allows separate control of the 2DEG density and potential modulation strength in GaAs heterostucture based artificial superlattices. Electrical transport measurements show clean Shubnikov de Haas oscillations at low magnetic fields and well quantised Hall plateaus, indicating the high-quality of the sample and showing that the fabrication process hardly introduces any disorder. By varying the biases applied to two gates, we also show that the periodic modulation potential can be changed from weak ($U\ll E_F$) to strong ($U\geqslant E_F$), and the 2DEG density can be easily tuned from $4\times10^{10} cm^{-2}$ to $1.6\times10^{11} cm^{-2}$.

\begin{figure}[!ht]
	\centering
\includegraphics[width=0.99\linewidth]{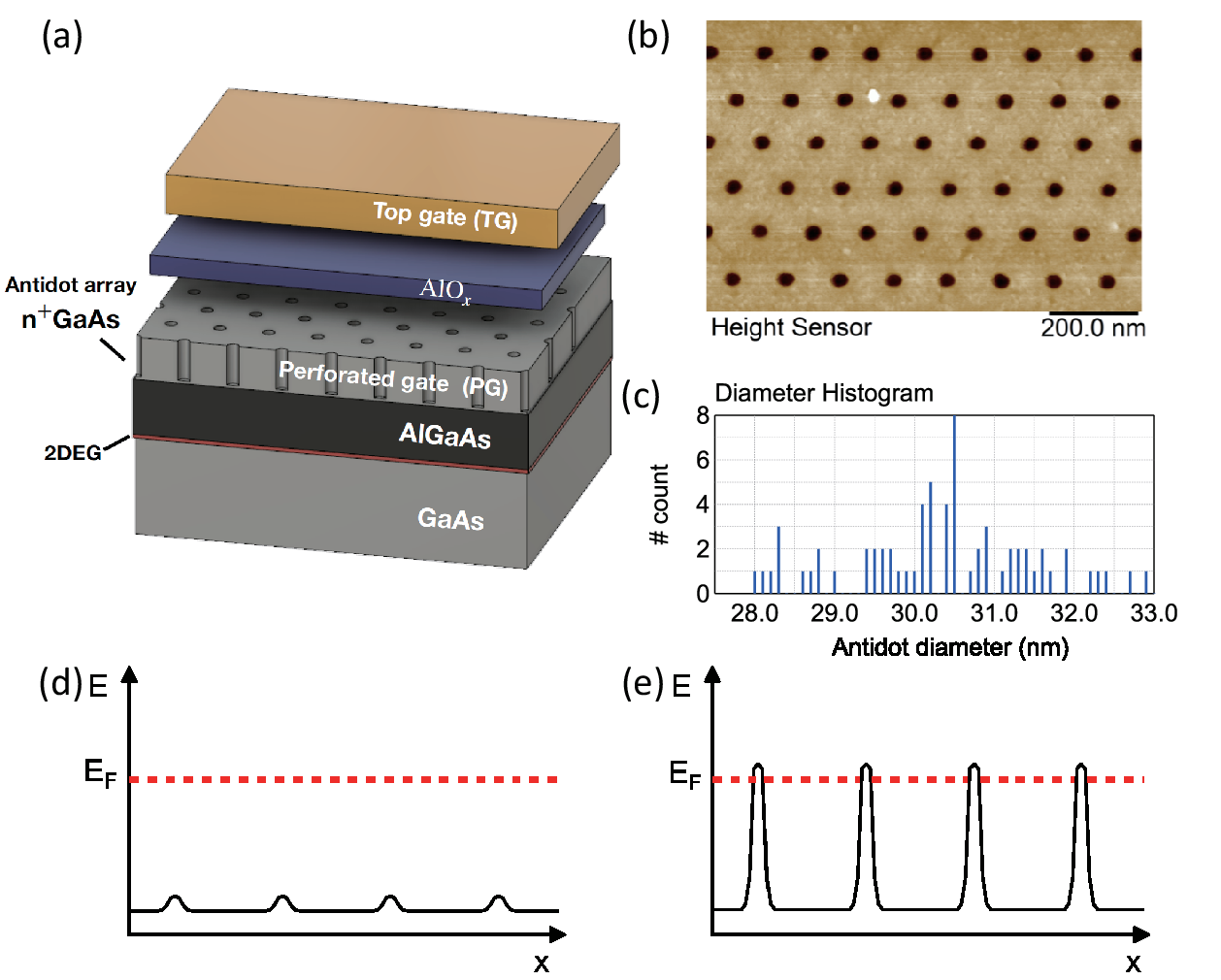}
\caption{\small (a) 3D schematic of the our device. The bottom perforated gate is a heavily doped GaAs layer grown in the MBE chamber. The wafer is then processed and etched with a periodic antidot array. (b) AFM image of a device showing the etched n$^+$GaAs capping layer with a triangular superlattice with a lattice constant of $120 nm$. (c) Statistical analysis of the etched antidot array showing the uniformity of the etched superlattice. The etched antidots have a average diameter of $30nm$ with a standard deviation of 1nm. (d) (e) Schematics demonstrating the two limits of the modulation potential achievable by adjusting the bias applied to the top gate: (d) the weak modulation regime where the modulation potential is much smaller than the Fermi energy $U\ll E_F$, and (d) the strong modulation regime where the modulation potential is comparable to or larger than the Fermi energy $U\geqslant E_F$.}
\label{fig:device}
\end{figure}

In our approach we use an accumulation-mode GaAs-AlGaAs heterostructure with a double-layer gate design as depicted by the schematic in Figure~\ref{fig:device}(a): the lower gate is a perforated gate (PG), which is an epitaxially grown layer of n$^+$GaAs etched with periodic hole arrays as shown in Figure~\ref{fig:device}(b). This heavily doped GaAs gate design has the advantage of allowing the gate to be very close to the 2DEG while minimising scattering from surface states, which is the dominant scattering process in shallow heterostructures~\cite{Wang13}, and thus improves the electron mobility. 
The wafer used in this study is a shallow undoped (100) GaAs/Al$_x$Ga$_{1-x}$As heterostructure comprising a $15nm$ $\delta$-doped n$^+$GaAs cap, a $10nm$ bulk-doped n$^+$GaAs cap, a $10nm$ undoped GaAs cap and a $50 nm$ Al$_x$Ga$_{1-x}$As layer on a GaAs buffer layer. A total of $25nm$ thick n$^+$GaAs cap serves as the in-situ perforated gate. The effective distance between the hetero-interface and the perforated gate is $60nm$.

To pattern the n$^+$GaAs gate with a superlattice we developed a low-power inductively coupled plasma-reactive ion etching (ICP-RIE) process to minimize damage to the heterostructure due to ion bombardment. Positive e-beam resist AR-P 6200 patterned with a Raith e-beam writer is utilized as the etch-mask for the ICP-RIE step. A low flow of etchant BCl$_3$ is used to control the etch depth, stopping in the n$^+$ gate layer so that only the conducting layer is removed. As shown by the Atomic Force Microscopy(AFM) image of an etched dot array in Fig~\ref{fig:device}(b), the patterned superlattice is in the form of a lateral 2D triangular antidot lattice with a lattice constant of 120nm. Such small superlattices are susceptible to any non-uniformity in the lithographic patterning. A statistical analysis of the AFM image shows that the resulting holes are highly periodic and uniform, with a diameter of 30nm consistent with the patterned resist.

In a normal operation of the device, a positive bias is applied to the perforated gate, which induces electrons at the GaAs-AlGaAs hetero-interface and controls the electron density. Insulated by a thin layer of aluminium oxide, the top layer is an overall metal top gate (TG). Since the top gate is screened by the lower perforated gate except where the perforated gate is drilled, i.e at the vertices of the triangular superlattice, it can be used to manipulate the strength of the periodic modulation potential $U$. Depending on the voltage applied to the top gate, the device can either operate in the weak-modulation regime $U\ll E_F$ as shown in Figure~\ref{fig:device}(d) or the strong-modulation regime $U\geqslant E_F$ as shown in Figure~\ref{fig:device}(e). When the top gate is positively or weakly negatively biased, 2DEG can still form underneath it at the vertices of the superlattice due to the positive bias on the perforated gate. In this case the modulation potential is quite weak and much smaller than $E_F$. When the top gate is strongly negatively biased, it fully depletes the electrons beneath so that no 2DEG exists at the vertices of the superlattice. This creates a very strong periodic modulation potential that is comparable to or larger than $E_F$.

\begin{figure}[!ht]
	\centering
\includegraphics[width=0.95\linewidth]{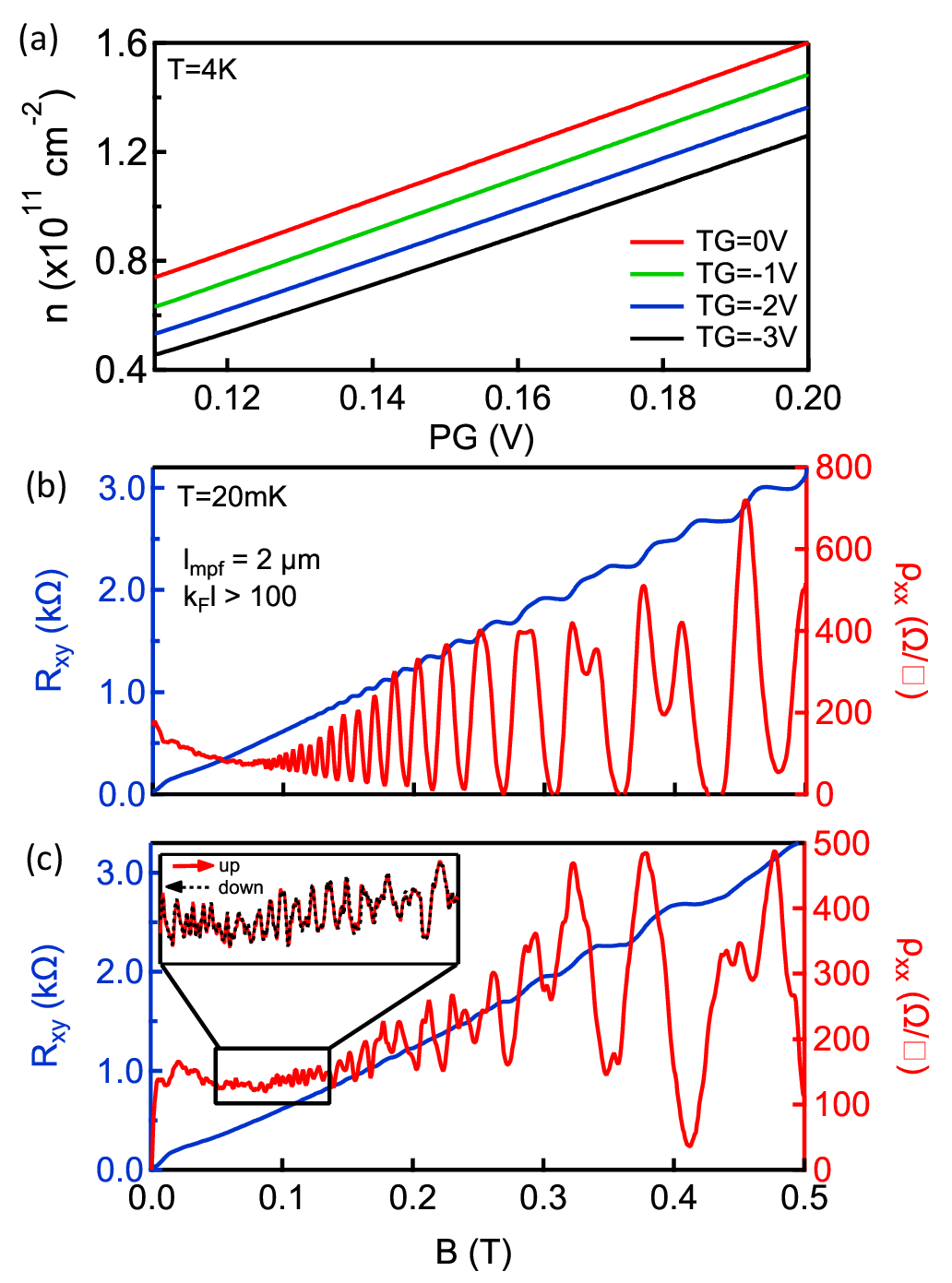}
\caption{\small (a) Density of the 2DEG as a function of the voltage applied to the perforated gate (PG) for different top gate (TG) voltages at $T=4K$ measured from the low-field Hall slope. (b) and (c) Longitudinal magneto-resistance and Hall resistance of the device at a 2DEG density of $n=1\times10^{11} cm^{-2}$ and $T=20mK$ for $V_{TG} = 0$ and $V_{TG} = -3V$ respectively. Inset of (c) shows a zoom-in of the boxed area (magneto-resistance between 0.06T and 0.15T) including both up and down field sweeps to show that the fine structures in the oscillations are highly reproducible.}
\label{fig:char}
\end{figure}

We first present characterization of the device at 4K using standard lock-in techniques~\cite{Chen12} with a small AC excitation voltage of $100\mu V$. Figure~\ref{fig:char}(a) shows the density of the device extracted from the low-field Hall slope as a function of $V_{PG}$ for different voltages on the top gate. The density of the device is a linear function of $V_{PG}$ with an average slope of $93\times10^{10}cm^{-2}/V$, which can be understood as the capacitance of a parallel-plate capacitor formed by the perforated gate and the 2DEG. 
The top gate has a much weaker effect on the 2DEG density, since it is mainly screened from the 2DEG by the perforated gate underneath it. From the shifting of the density traces with applied $V_{TG}$ in Figure~\ref{fig:char}(a), the effect of the $V_{TG}$ on 2DEG density is calculated to be $1.1\times10^{10} cm^{-2}/V$, more than 80 times smaller than that of $V_{PG}$. 

In Figure~\ref{fig:char}(b) and (c) we show the longitudinal magneto-resistance and Hall resistance measured at dilution refrigerator base temperature $T=20mK$ and electron density $n=1\times10^{11} cm^{-2}$ for $V_{TG}=0$ and $V_{TG}=-3V$. At $V_{TG}=0$, the device shows low resistivity, clean Shubnikov de Haas oscillations and Hall quantization at magnetic fields as low as 0.1T, comparable to a 2DEG in the same wafer without superlattice patterning. This suggests that the heterostructure maintained high-quality and there is low damage to the heterostructure caused by the fabrication process, especially from the etching of the superlattice. From the resistance we estimate the mean free path in this device is around $2 \mu m$ and $k_F l$ is larger than 100 at $n=1\times10^{11} cm^{-2}$. 
This mean free path is much larger than the lattice spacing of 120nm, which guarantees the electrons ``feel" the artificial lattice potential before they scatter. 
Biasing $V_{TG}$ to $-3V$ turns on the superlattice potential, and the magneto-resistance is strongly modified with a complicated beating pattern. As shown in the inset of Figure~\ref{fig:char}(c), the oscillations in the magneto-resistance are highly reproducible even at magnetic fields down to 60mT. This indicates that the complex magneto-resistance is not due to noise but must be caused by the periodic potential landscape imposed by the patterned gate.

\begin{figure}[!ht]
	\centering
\includegraphics[width=0.99\linewidth]{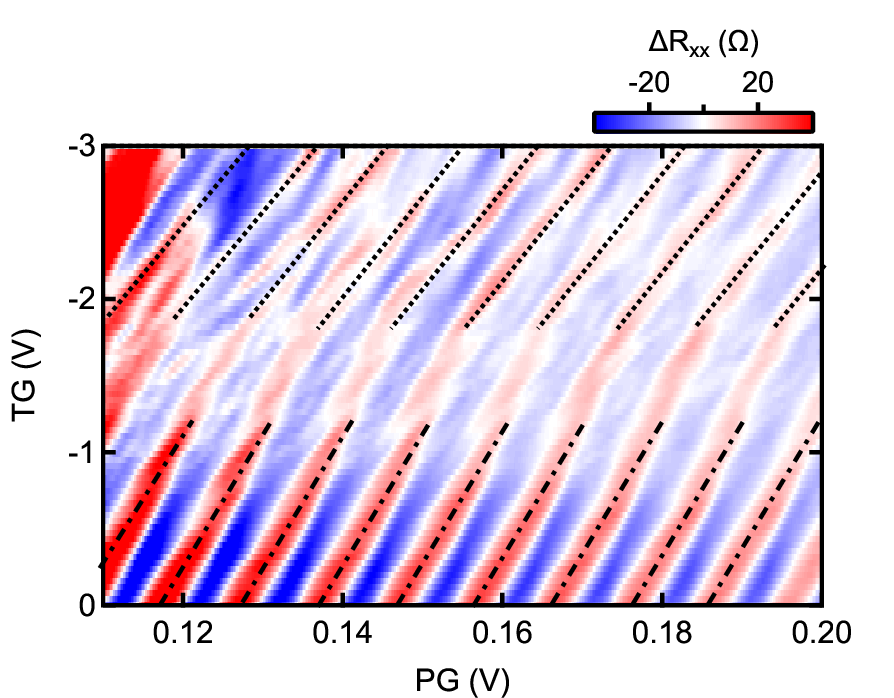}
\caption{\small  Background removed magneto-resistance of the device at a fixed magnetic field $B=0.2T$ plotted as a function of $V_{PG}$ and $V_{TG}$ voltages. The approximate number of electrons per unit cell is varied from 3 ($V_{PG}=0.11V$, $V_{TG}=-3V$) in the top left corner to 10 ($V_{PG}=0.2V$, $V_{TG}=0V$) in the bottom right corner. Dash-dotted and dotted lines are guides to the eye indicating the change of slope for the low $V_{TG}$ ($V_{TG}<-1V$) and high $V_{TG}$ ($V_{TG}>-2V$) regimes. }
\label{fig:map}
\end{figure}

To estimate the strength of the modulation potential and investigate how it is varied by $V_{TG}$, we plot in Figure~\ref{fig:map} the background-removed magneto-resistance~\cite{supp} at $B=0.2T$ as a function of the voltages applied to the perforated gate and the top gate. Over the entire map, the number of electrons per unit cell is varied from approximately 3 to 10 with the highest number of electrons per unit cell at $V_{PG}=0.2V$, $V_{TG}=0$ in the top left corner and the lowest number of electrons per unit cell at $V_{PG}=0.11V$, $V_{TG}=-3V$ in the bottom right corner. The colour plot shows a series of blue and red stripes which correspond to magneto-resistance oscillations as the density of the 2DEG is varied at $B=0.2T$. The tilt of the stripes towards the right is caused by the weak effect of $V_{TG}$ on the 2DEG density. Therefore, the slope of the stripes is a measure of the relative strength of the capacitive coupling between TG-2DEG and PG-2DEG. From the dash-dotted lines, this slope is estimated around $85$, in good agreement with the $4K$ data shown in Figure~\ref{fig:char}(a).

An important feature observed in Figure~\ref{fig:map}, highlighted by the dash-dotted and dotted lines, is that the slope of the stripes undergoes a clear change from $\sim85$ to $\sim65$ as the top gate is biased more negatively. This can be understood as follows. For the the low $V_{TG}$ regime ($V_{TG}<-1V$), the modulation potential is weak and much smaller than the Fermi energy ($E_F$) as depicted in Figure~\ref{fig:device}(d). In this case, electrons exist everywhere in the system including underneath the top gate, i.e, at the vertices of the superlattice. For the high $V_{TG}$ regime ($V_{TG}>-2V$), the top gate fully depletes the electrons beneath it and created a potential barrier comparable to or larger than $E_F$. This results in a strong modulation potential as depicted in Figure~\ref{fig:device}(e) and a decrease in the effective area of the perforated gate, which reduces the capacitive coupling between the perforated gate and the 2DEG and thus the slope of the stripes. Therefore, this change in the slope is a clear indication that the modulation potential $U$ can be effectively tuned from the weak ($U\ll E_F$) to strong ($U\geqslant E_F$) regimes by simply varying $V_{TG}$.

In conclusion, we fabricated a high-quality electron system with a superimposed small-period triangular superlattice. We used shallow GaAs heterostructures with an in-situ gate and a double-layer design. Magneto-transport measurements of the system show low resistance, clean oscillations and well quantised Hall plateaus at very small magnetic fields, which indicates the fabrication process caused minimal damage to the heterostructure.
By changing the voltage applied to two separate gates, we demonstrated that density and modulation strength can be tuned separately in these devices over a wide range. This device architecture can serve as a flexible platform for future band structure engineering experiments.

\acknowledgements
We acknowledge useful discussions with O. P. Sushkov and Z. E. Krix. This work is supported by the Australian Research Council Centre of Excellence in Future Low-Energy Electronics Technologies (CE170100039). A. D. W. acknowledges gratefully support of DFG-TRR160, BMBF - Q.Link.X 16KIS0867, and the DFH/UFA CDFA-05-06. All devices were fabricated using facilities at the NSW and ACT Nodes of the Australian National Fabrication Facility. AFM images were taken using the facilities of Microscopy Australia at the Electron Microscope Unit within the Mark Wainwright Analytical Centre at UNSW Sydney.

\section*{Data Availability Statement}
The data that support the findings of this study are available from the corresponding author upon reasonable request.

\bibliographystyle{unsrt}

\begin{thebibliography}{99}
\bibitem{supp} Original magneto-resistance is smoothed to remove any oscillations and then subtracted from the original trace as the background.
\end{thebibliography}

\end{document}